\documentclass[12pt]{article}
\usepackage{graphicx}
\usepackage{epsfig}
\usepackage{axodraw}
\usepackage{amssymb}

\catcode`@=11
\newcount\@tempcntc
\def\@citex[#1]#2{\if@filesw\immediate\write\@auxout{\string\citation{#2}}\fi
  \@tempcnta\z@\@tempcntb\m@ne\def\@citea{}\@cite{\@for\@citeb:=#2\do
    {\@ifundefined
       {b@\@citeb}{\@citeo\@tempcntb\m@ne\@citea\def\@citea{,}{\bf ?}\@warning
       {Citation `\@citeb' on page \thepage \space undefined}}%
    {\setbox\z@\hbox{\global\@tempcntc0\csname b@\@citeb\endcsname\relax}%
     \ifnum\@tempcntc=\z@ \@citeo\@tempcntb\m@ne
       \@citea\def\@citea{,}\hbox{\csname b@\@citeb\endcsname}%
     \else
      \advance\@tempcntb\@ne
      \ifnum\@tempcntb=\@tempcntc
      \else\advance\@tempcntb\m@ne\@citeo
      \@tempcnta\@tempcntc\@tempcntb\@tempcntc\fi\fi}}\@citeo}{#1}}
\def\@citeo{\ifnum\@tempcnta>\@tempcntb\else\@citea\def\@citea{,}%
  \ifnum\@tempcnta=\@tempcntb\the\@tempcnta\else
   {\advance\@tempcnta\@ne\ifnum\@tempcnta=\@tempcntb \else \def\@citea{--}\fi
    \advance\@tempcnta\m@ne\the\@tempcnta\@citea\the\@tempcntb}\fi\fi}
\catcode`@=12
\newcommand{\im}{\mathop{\mathrm{Im}}\nolimits}
\newcommand{\Li}{\mathop{\mathrm{Li}}\nolimits}

\begin{document}

\title{
\vskip-3cm{\baselineskip14pt
\centerline{\normalsize DESY 06-073\hfill ISSN 0418-9833}
\centerline{\normalsize hep-ph/0607201\hfill}
\centerline{\normalsize July 2006\hfill}}
\vskip1.5cm
Heavy-quark QCD vacuum polarisation function:
analytical results at four loops
}
\author{Bernd A.~Kniehl, Anatoly V.~Kotikov\thanks{Permanent address:
Bogoliubov Laboratory of Theoretical Physics,
Joint Institute for Nuclear Research, 141980, Dubna (Moscow region), Russia.}\\
{\normalsize\it II Institut f\"ur Theoretische Physik,
Universit\"at Hamburg},\\
{\normalsize\it Luruper Chaussee 149, 22761 Hamburg, Germany}}

\date{5 June 2006}

\maketitle

\begin{abstract}
Fully analytical results for the first two moments of the heavy-quark vacuum
polarisation function at four loops in quantum chromo-dynamics are presented.

\end{abstract}

\newpage

\section{Introduction}

During the past several years, there has been significant progress in the
determination of the values of the strong-coupling constant $\alpha_s$ at the
$Z$-boson mass scale and the heavy-quark masses $m_h$ in perturbative
quantum chromo-dynamics (QCD).
The matching relations which determine the finite discontinuities of
$\alpha_s$ at the heavy-quark thresholds in the modified
minimal-subtraction ($\overline{\mathrm{MS}}$) scheme and are necessary for
a precise determination of $\alpha_s$ from a global data fit were recently
obtained in semi-analytical \cite{ChKSt} and analytical \cite{KKOV06} form at
the four-loop level.
Furthermore, there is significant progress in multi-loop technology aiming
towards an analytical evaluation of the QCD $\beta$ function at five loops.

In order to precisely determine heavy-quark masses with the help of QCD sum
rules \cite{Novikov}, it is necessary to know in detail the heavy-quark
contribution to the QCD vacuum polarisation function.
The most important ingredients for this analysis are the lowest moments of the
polarisation function (see Ref.~\cite{Shifman}).
Indeed, it was found that the first moment is best suited for this analysis,
since it has the weakest dependence on non-perturbative effects and the
details of the threshold region, thus leading to the smallest theoretical
uncertainty \cite{Kuhn}.

At the four-loop level, the study of the first two moments has been started
in Ref.~\cite{CKMS} with calculations of diagrams including two internal loops
of massive and massless quarks coupled to gluons.
The results in Ref.~\cite{CKMS} contain two numerical constants, denoted as
$N_{10}$ and $N_{20}$, which were not expressed in terms of basic
transcendental numbers.
In the meantime, the corresponding Feynman diagrams have been calculated
analytically \cite{Schroder,KnKo05}.
The terms proportional to $\alpha_s^j n_l^{j-1}$, where $n_l$ in the number of
light quarks, are known in all orders $j$ of perturbative QCD \cite{Grozin}.
References~\cite{ChKSt06,Czakon} contain the complete four-loop contributions
to the first two Taylor coefficients from non-singlet diagrams.
The singlet contributions were studied in Refs.~\cite{Czakon,Groote}. 

The results of Refs.~\cite{ChKSt06,Czakon} contain some four-loop tadpoles
which are only given in numerical form.
The purpose of this Letter is to evaluate them in terms of standard
transcendental numbers so as to represent the first two moments of the
heavy-quark vacuum polarisation function at four loops in completely
analytical form.

The content of this Letter is as follows.
Section~\ref{sec:two} contains the basic formulae.
In Section~\ref{sec:three}, we present our analytic results for the tadpoles
that, in Refs.~\cite{ChKSt06,ChFStT}, are called $T_{54}$, $T_{62}$, and
$T_{91}$ as well as for the first two moments of the heavy-quark vacuum
polarisation at four loops in QCD.
A brief summary is given in Section~\ref{sec:four}.

\section{Approach}
\label{sec:two}

Since the vector current $j^{\mu}(x) = \overline{\Psi}(x)\gamma^{\mu} \Psi(x)$
constructed from the heavy-quark field $\Psi(x)$ is conserved, the correlator,
\begin{eqnarray}
\Pi^{\mu\nu}(q)&=&i \int dx e^{iqx}\langle 0|T j^{\mu}(x) j^{\nu}(0) |0\rangle
\nonumber\\
&=& (-q^2 g^{\mu\nu}+q^{\mu}q^{\nu})\Pi (q^2),
\end{eqnarray}
can be expressed in terms of a single scalar function $\Pi(q^2)$.
The latter is of great phenomenological interest because it is related to the
experimental observable
\begin{eqnarray}
R(s)&=&\frac{\sigma(e^+e^- \to \mbox{hadrons})}{\sigma(e^+e^- \to \mu^+\mu^-)}
\nonumber\\
&=& 12 \pi\im \Pi(s+i\epsilon), 
\label{eq:r}
\end{eqnarray}
where $s$ is the square of the $e^+e^-$ centre-of-mass energy.
Equation~(\ref{eq:r}) is equivalent to an infinite number of equalities
between experimental moments,
\begin{equation}
M^{\rm exp}_n = \int \frac{ds}{s^{n+1}}R(s),
\end{equation}
and their theoretical counterparts
\begin{equation}
M^{\rm th}_n = \frac{9e_h^2}{4} \left(\frac{1}{4\overline{m}_h^2}\right)^n
\overline{C}_n ,
\end{equation}
where $e_h$ is the electric-charge quantum number of the heavy quark,
$\overline{m}_h=m_h\left(\overline{m}_h\right)$, and
$\overline{C}_n$ are the coefficients of the Taylor expansion
\begin{equation}
\Pi(q^2) = \frac{3e_h^2}{16\pi^2} \sum_{n\geq0}\overline{C}_n 
\overline{z}_n
\end{equation}
in $\overline{z}=q^2/(4\overline{m}_h^2)$.
The perturbative calculation of
\begin{equation}
\overline{C}_n =  \overline{C}_n^{(0)} 
+ \frac{\alpha_s}{\pi} \overline{C}_n^{(1)}
+ {\left(\frac{\alpha_s}{\pi}\right)}^2 \overline{C}_n^{(2)}
+ {\left(\frac{\alpha_s}{\pi}\right)}^3 \overline{C}_n^{(3)}
+ \cdots,
\end{equation}
where $\alpha_s=\alpha_s\left(\overline{m}_h^2\right)$, only involves tadpole
diagrams and is, therefore, considerably simpler than the one of
$\Pi(q^2)$, which has a complicated dependence on $\overline{z}$.
Here, it is understood that $\alpha_s(\mu)$ and $m_h(\mu)$ are defined in the
$\overline{\mathrm{MS}}$ renormalisation scheme, which is based on dimensional
regularisation with $D=4-2\epsilon$ space-time dimensions and 't~Hooft mass
scale $\mu$.

Thanks to the strong hierarchy among the quark masses, one may split the
number $n_f=n_l+n_h$ of quark flavours into $n_h=1$ massive ones and $n_l$
massless ones.
In the following, we keep the variable $n_h$ generic.

\section{Results}
\label{sec:three}

In our previous paper~\cite{KnKo05}, we introduced a technique that allows one
to analytically evaluate a large class of four-loop tadpole diagrams with one
non-zero mass.
Specifically, the considered diagrams are transformed into integral
representations whose integrands contain only one-loop tadpoles with new
propagators having masses that depend on the variables of integration.
This technique is based on the differential equation method \cite{DEM}.
A similar technique was applied to certain types of two-loop diagrams in
Ref.~\cite{FKVpl}.

To illustrate the usefulness of this technique, in Ref.~\cite{KnKo05}, we also
evaluated the tadpoles that are denoted in Ref.~\cite{CKMS} as $N_{10}$ and
$N_{20}$.
The finite parts of our results agree with the numerical results of
Ref.~\cite{CKMS} and the analytical ones of Ref.~\cite{Schroder}, while our
analytical results for the $O(\epsilon)$ terms are new.

\begin{figure}
\begin{eqnarray}
\hspace{2cm}
\mbox{{
%\begin{picture}(50,30)(0,0)
\begin{picture}(50,70)(0,0)
%\DashCArc(40,0)(35,120,240){3}
\CArc(40,0)(35,120,240)
%%\GlueArc(20,15)(15,90,270){1}{6}
%\GlueArc(20,15)(15,90,270){1}{6}
%\DashCArc(0,0)(35,-60,60){3}
\CArc(0,0)(35,-60,60)
%\CArc(20,-15)(15,90,270)
%\CArc(20,-15)(15,-90,90)
\CArc(20,0)(30,90,270)
\CArc(20,0)(30,-90,90)
\Vertex(20,30){2}
%\Vertex(20,0){2}
\Vertex(20,-30){2}
%\Line(20,0)(17,-5)
\DashLine(20,30)(20,-30){3}
%\Line(20,0)(23,-5)
\Text(20,-50)[c]{$T_{54}$} 
%\Text(3,18)[r]{$\scriptstyle\alpha_1$} 
%\Text(38,18)[l]{$\scriptstyle\alpha_2$} 
\end{picture}
}} ~~~~~~~~~~~
%\mbox{{
%%\begin{picture}(50,30)(0,0)
%\begin{picture}(50,70)(0,0)
%\CArc(20,15)(15,90,270)
%%%\GlueArc(20,15)(15,90,270){1}{6}
%%\GlueArc(20,15)(15,90,270){1}{6}
%\CArc(20,15)(15,-90,90)
%\CArc(20,-15)(15,90,270)
%\CArc(20,-15)(15,-90,90)
%\CArc(20,0)(30,90,270)
%\CArc(20,0)(30,-90,90)
%\Vertex(20,30){2}
%\Vertex(20,0){2}
%\Vertex(20,-30){2}
%%\Line(20,0)(17,-5)
%%\Line(20,0)(23,-5)
%\Text(20,-50)[c]{$T_{61}$} 
%%\Text(3,18)[r]{$\scriptstyle\alpha_1$} 
%%\Text(38,18)[l]{$\scriptstyle\alpha_2$} 
%\end{picture}
%}} ~~~~~~~~~~~
\mbox{{
%\begin{picture}(50,30)(0,0)
\begin{picture}(50,70)(0,0)
\DashCArc(20,15)(15,90,270){3}
%%\GlueArc(20,15)(15,90,270){1}{6}
%\GlueArc(20,15)(15,90,270){1}{6}
\DashCArc(20,15)(15,-90,90){3}
\CArc(20,-15)(15,90,270)
\CArc(20,-15)(15,-90,90)
\CArc(20,0)(30,90,270)
\CArc(20,0)(30,-90,90)
\Vertex(20,30){2}
\Vertex(20,0){2}
\Vertex(20,-30){2}
%\Line(20,0)(17,-5)
%\Line(20,0)(23,-5)
\Text(20,-50)[c]{$T_{62}$} 
%\Text(3,18)[r]{$\scriptstyle\alpha_1$} 
%\Text(38,18)[l]{$\scriptstyle\alpha_2$} 
\end{picture}
}} ~~~~~~~~~~~
\mbox{{
%\begin{picture}(50,30)(0,0)
\begin{picture}(50,70)(0,0)
\CArc(20,0)(30,90,270)
\CArc(20,0)(30,-90,90)
\Vertex(20,30){2}
%\Vertex(20,0){2}
\Vertex(20,-30){2}
\Vertex(40,20){2}
\Vertex(40,-20){2}
\Vertex(0,-20){2}
\Vertex(0,20){2}
%\Line(20,0)(17,-5)
\DashLine(20,30)(20,-30){3}
\DashLine(40,20)(0,-20){3}
\DashLine(40,-20)(0,20){3}
%\Line(20,0)(23,-5)
\Text(20,-50)[c]{$T_{91}$} 
%\Text(3,18)[r]{$\scriptstyle\alpha_1$} 
%\Text(38,18)[l]{$\scriptstyle\alpha_2$} 
\end{picture}
}}
\nonumber
\end{eqnarray}
\vspace{1cm}
\caption{\label{fig:tad}Massive four-loop tadpole diagrams $T_{54}$, $T_{62}$,
and $T_{91}$.}
\end{figure}
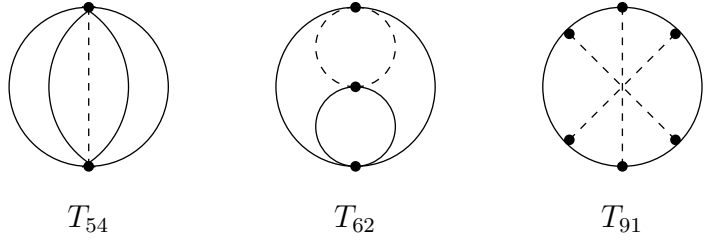

These results together with those from Ref.~\cite{KKOV06} offer us the
opportunity to obtain analytical results for terms of the expansion in
$\epsilon$ of the master integrals denoted as $T_{54}$, $T_{62}$, and $T_{91}$
in recent papers \cite{ChKSt06,ChFStT} for which only numerical values have
been available so far.
The corresponding Feynman diagrams are depicted in Fig.~\ref{fig:tad}.
Specifically, $T_{62}$ can be extracted from $N_{10}$ \cite{KnKo05} using 
Eqs.~(16) and (18) of Ref.~\cite{CKMS}.
This result can then be used in combination with $T_{91}$ \cite{KKOV06} to
extract $T_{54}$ from Eq.~(50) of Ref.~\cite{ChFStT}.
The results read
\begin{eqnarray}
\frac{T_{54}}{N} &=& -\frac{1}{\epsilon^4} - \frac{9}{2\epsilon^3} 
-\frac{415}{36\epsilon^2}
-\frac{4991}{216\epsilon} -\frac{89383}{1296} 
+ \epsilon \left( -\frac{3367679}{7776} +  \frac{1792}{9} \zeta(3) \right)
\nonumber\\
&&{}+ \epsilon^2 \left(-\frac{137735095}{46656} 
-\frac{4352\pi^4}{135} + \frac{68992}{27} \zeta(3) 
+  \frac{1024}{3} b_4\right) 
\nonumber\\
&&{}+  \epsilon^3 \left(
-\frac{4908181487}{279936} - \frac{167552\pi^4}{405} + \frac{17408\pi^4}{45}
\ln 2 + \frac{1602496}{81} \zeta(3)  \right.
\nonumber \\
&&{}-\left. \frac{87296}{3} \zeta(5) + \frac{39424}{9} b_4+ 4096 b_5 
\right) 
+ O(\epsilon^4),
\label{T54}\\
\frac{T_{62}}{N} &=& 
\frac{3}{2\epsilon^4} + \frac{4}{\epsilon^3} +\frac{38}{\epsilon^2}
+\frac{1}{\epsilon}\left(\frac{44}{3} + \frac{16}{3} \zeta(3)\right)
-118-\frac{4\pi^4}{15}+ 88\zeta(3)
\nonumber \\ 
&&{}+ \epsilon \left( -1156 
- \frac{374\pi^4}{45}  
+ \frac{2152}{3} \zeta(3) 
+ 96 \zeta(5) + 64 b_4\right)
\nonumber \\
&&{}+ \epsilon^2 \left(
-\frac{20938}{3} - \frac{710\pi^4}{9} 
 + \frac{2416\pi^4}{45} \ln 2 - \frac{8\pi^6}{21}
+\frac{12952}{3} \zeta(3)
\right.
\nonumber\\
&&{}+\left. 
 \frac{64}{3} \zeta^2(3) 
- 2400 \zeta(5) + 704 b_4 + 512 b_5 \right) + O(\epsilon^3),
\label{T62}\\
T_{91} &=&  -\frac{53\pi^4}{15} \ln 2 +\frac{873}{2} \zeta(5) - 48 b_5
+ O(\epsilon),
\label{T91}
\end{eqnarray}
where 
\begin{eqnarray}
b_4 &=&-\frac{1}{3}(\pi^2-\ln^2 2)\ln^2 2+8\Li_4\left(\frac{1}{2}\right),
\nonumber\\
b_5 &=&\frac{1}{45}(5\pi^2-3\ln^2 2)\ln^3 2+8\Li_5\left(\frac{1}{2}\right),
\label{b}
\end{eqnarray}
and $N=\Gamma^4(1+\epsilon)(m^2)^{-4\epsilon}$ is a normalisation factor.
The choice $N=1$ corresponds to the normalisation adopted in
Ref.~\cite{ChFStT}.
The terms through $O(\epsilon^2)$ in Eq.~(\ref{T54}) and those through
$O(\epsilon)$ in Eq.~(\ref{T62}) agree with Eqs.~(38) and (42) in
Ref.~\cite{ChFStT}, respectively, while the other terms represent new results.

Equations~(\ref{T54})--(\ref{T91}) extend the set of four-loop tadpole
master integrals with one non-vanishing mass known in fully analytical form,
and represent essential ingredients for high-precision predictions of numerous
observables of current phenomenological interest.
After changing the overall normalisation, we find agreement with the numerical
values given in Ref.~\cite{Schroder}.
We note in passing that also the other master integrals presented in
Ref.~\cite{Schroder} can be written in a more compact form if the combinations
$b_4$ and $b_5$ of Eq.~(\ref{b}) are introduced.

Equation~(\ref{T91}) is the key result of Ref.~\cite{KKOV06} and a crucial
ingredient for the four-loop matching condition of $\alpha_s(\mu)$ at the
heavy-quark thresholds \cite{ChKSt,KKOV06}.
Equations~(\ref{T54})--(\ref{T91}) are the missing links for a fully
analytical representation of the moments $\overline{C}_0^{(3)}$ and
$\overline{C}_1^{(3)}$ \cite{ChKSt06}, which is the main goal of this paper.

Our results for $\overline{C}_0^{(3)}$ and $\overline{C}_1^{(3)}$ read
\begin{eqnarray}
\overline{C}_{0}^{(3)} &=& 
n_l^2 \left(\frac{17897}{69984}- \frac{31}{162} \zeta(3) \right)
+n_ln_h \left(
\frac{7043}{34992} + \frac{49\pi^4}{12960}\right.
\nonumber\\
&&{}-\left.\frac{127}{324} \zeta(3)- \frac{1}{36}b_4
\right)
+ n_h^2 \left(\frac{610843}{2449440}- \frac{661}{2835} \zeta(3) \right)
\nonumber\\
&&{}+ n_l \left(
-\frac{71629}{46656} + \frac{8533\pi^4}{116640} 
- \frac{21343}{3888} \zeta(3)- \frac{25}{324}b_4 
\right)
\nonumber\\
&&{}+ n_h \left(
-\frac{83433703}{8164800} + \frac{14873\pi^4}{18225}  
- \frac{14509529}{340200} \zeta(3) + \frac{5}{3} \zeta(5)\right.
\nonumber \\
&&{}-\left.\frac{7091}{810}b_4\right)
-\frac{5266559}{466560} - \frac{702959\pi^4}{699840} 
- \frac{1289\pi^4}{3645}\ln 2 
\nonumber\\
&&{}+  \frac{1688407}{12960} \zeta(3) - \frac{89}{24} \zeta(5)
-  \frac{313}{1944} b_4 -  \frac{328}{81} b_5,
\nonumber\\
\overline{C}_{1}^{(3)} &=& 
n_l^2 \left(\frac{42173}{98415}- \frac{112}{405} \zeta(3) \right)
+n_ln_h \left(
\frac{262877}{787320} + \frac{1421\pi^4}{174960}\right.
\nonumber\\
&&{}-\left.\frac{38909}{58320} \zeta(3)- \frac{29}{486}b_4\right)
+ n_h^2 \left(\frac{163868}{295245}- \frac{3287}{7290} \zeta(3) \right)
\nonumber\\
&&{}+ n_l \left(
-\frac{9338899}{2099520} + \frac{372689\pi^4}{839808}  
- \frac{48350497}{1399680} \zeta(3) - \frac{4793}{58320}b_4
\right)
\nonumber\\
&&{}+ n_h \left(
-\frac{27670774337}{1414551600} + \frac{1447057\pi^4}{765450} 
-\frac{95617883401}{943034400} \zeta(3)
\right.
\nonumber \\
&&{}+\left.\frac{128}{27} \zeta(5)
-\frac{348701}{17010}b_4\right)
-\frac{5397779543}{146966400} - \frac{2653167371\pi^4}{881798400}
\nonumber\\
&&{}-\frac{359687\pi^4}{229635}\ln 2 
 + \frac{17554601717}{32659200} \zeta(3)
-\frac{3655}{10206} \zeta(5)
-\frac{84951877}{2449440} b_4
\nonumber\\
&&{}- \frac{127480}{5103} b_5.
\end{eqnarray}
These results agree with the semi-analytical ones presented in
Refs.~\cite{ChKSt06,Czakon}. 

As the authors of Refs.~\cite{ChKSt06,Czakon}, we conclude that the inclusion
of the new four-loop corrections changes the actual values of the charm- and
bottom-quark masses derived from QCD sum rules only very little, while it
strongly decreases their theoretical uncertainties: approximately by factors
of three and four in the case of charm and bottom, respectively.

\section{Conclusion}
\label{sec:four}
 
We presented analytical results for the four-loop tadpole master integrals
involving one non-vanishing mass that are called $T_{54}$, $T_{62}$, and
$T_{91}$ in the recent literature \cite{ChKSt06,ChFStT}, for which only
numerical values had been available so far.
They represent essential ingredients for high-precision predictions of a number
of observables of current phenomenological interest.
As such an application, we considered the vacuum polarisation induced by a
heavy quark in the four-loop approximation of QCD and expressed the first two
coefficients of its Taylor expansion in the ratio of virtuality to
heavy-quark mass in fully analytical form.
The introduction of the characteristic combinations $b_4$ and $b_5$ of basic
transcendental numbers turned out the be useful in order to compactify the
expressions.
Other applications include the four-loop QCD correction to the electroweak
$\rho$ parameter, which has just been presented in semi-analytic form
\cite{ChetyrkinRho}.

\section*{Acknowledgements}

We are grateful to O.I. Onishchenko for useful communications and technical
advice related to the programming language {\tt Mathematica}, and to O.L.
Veretin for checking some of our formulae and for technical advice related to
the PSLQ program \cite{PSLQ}, which helped us in checking the $O(\epsilon)$
terms of the four-loop tadpoles $N_{10}$ and $N_{20}$.
A.V.K. was supported in part by the RFBR Foundation through Grant No.\
05-02-17645-a and the Heisenberg-Landau-Programm.
This work was supported in part by BMBF Grant No.\ 05 HT4GUA/4 and HGF Grant
No.\ NG-VH-008.

\end{document}